# Identify the stiffness of DNA via deep learning

Haiqian Yang[1, 2*], Liu Yang[3*], Shaobao Liu[1, 2†]

[1] *State Key Laboratory of Mechanics and Control of Mechanical Structures, Nanjing University of Aeronautics and Astronautics, Nanjing 210016, P.R. China*

[2] *Bioinspired Engineering and Biomechanics Center (BEBC), Xi'an Jiaotong University, Xi'an 710049, P.R. China*

[3] *School of Electronic and Information Engineering, Xi'an Jiaotong University, Xi'an 710049, P.R. China*

*\* Co-first authors: they contributed equally to this work*

*† Corresponding authors: sbliu@nuaa.edu.cn*

**Abstract** DNA detection is of great significance in the point-of-care diagnostics. The stiffness of DNA, varying with its sequence and mechanochemical environment, could be a potential marker for DNA identification. The steric configurations of DNA fragments with different stiffness were simulated by employing the Kirchhoff theory of thin elastic rods. We identified the stiffness of DNA with the trained convolutional neural network on the simulated image set. The identification accuracy reached 99.85%. The stiffness-based identification provided a promising approach for DNA detection.

# 1 Introduction

DNA carries the genomic information of human disease (Cummings et al. 2017; Kremer et al. 2017) or pathogenetic virus (Kuhn et al. 2013; Ou et al. 1988). Thus, DNA detection is of great significance in the point-of-care diagnostics (Gootenberg et al. 2017; Park et al. 2002). Currently, DNA detection relies on gold nanoparticle to change the local conductivity (Park, Taton and Mirkin 2002) or CRISP-Cas13a/C2c2 to target specific sequences (Gootenberg, Abudayyeh, Lee, Essletzbichler, Dy, Joung, Verdine, Donghia, Daringer and Freije 2017). A distinguishable and sensitive marker is of urgent need in DNA-based diagnostic detection (Boynton et al. 2003; Moinova et al. 2018). Emerging deep learning has become a promising method to dig through the genomic information from DNA sequence and make predictions with the learned features (Angermueller et al. 2016). Using deep learning, how the sequence specificities related to biological regulation and disease could be identified (Alipanahi et al. 2015; Bair and Tibshirani 2003; Kelley et al. 2016).

Mechano-genomics studies how the nucleus interacts with the local mechanochemical micro environment and how it regulates genomic programs and the maintenance of genomic integrity (Uhler and Shivashankar 2017). The stiffness of DNA varies with its sequence and mechanochemical environment, which could be a potential marker for DNA identification. For super-coiled duplex DNAs, their steric configurations are closely determined by their stiffness, which is linearly related to their persistent length (Benham 2010). The persistent length of such a worm like structure is highly related to its net charge and mechanochemical micro environment (Skolnick and Fixman 2010), which means DNAs may have varying stiffness among species (*e.g.*, human, virus, mutant). Besides, postreplication modifications (*e.g.* cytosine methylation) could change the stiffness of DNAs (Cassina et al. 2016). Discrete and continuum mechanical models have been developed to simulate the steric configuration of DNA. The site juxtaposition of DNA were simulated by Brownian dynamics (Huang et al. 2001; Jian et al. 1998). While the high local flexibility of short DNAs should be better described by anisotropic heterogeneous rod (Travers 2004) and could possibly be explained by dynamic features of Watson-Crick base pairs (Yuan et al. 2008), isotropic homogeneous elastic rod (*i.e.* Kirchhoff theory) is sufficient enough to coarse-grain long DNAs and make accurate predictions. The observed two orders of super-helicity have been successfully explained by analyzing DNA with Kirchhoff elastic rod theory (Benham 1977).

Also treating the DNA as an elastic rod, the worm-like chain model (Shimada and Yamakawa 1984) has been an effective model in describing long DNA (Peters and Maher 2010) and is applied to measure the stiffness of DNA (Cassina, Manghi, Salerno, Tempestini, Iadarola, Nardo, Brioschi and Mantegazza 2016).

Combing DNA mechanics and deep learning, the steric configurations of DNA fragments with different stiffness were simulated by employing the Kirchhoff theory of thin elastic rods to train a convolutional neural network (CNN). We identified the stiffness of DNAs in another simulated image set with the trained CNN.

**2 Materials and Methods**

Mechanical simulation provided an approach to generate clean and pure image sets to train the artificial intelligent model. To illustrate the process, we only generated two image sets ($2 \times 8000$ images, see supplementary material for some of them) of DNA fragments with the classical Kirchhoff elastic rod theory, namely the stiff DNA fragments ($224 \times 224$ pixels) (**Fig.1 A**) and the soft DNA fragments ($224 \times 224$ pixels) (**Fig.1 B**).

2.1 Statics of DNA fragments

The Kirchhoff elastic rod theory was widely used to describe the deformation of supercoiled DNA (Benham 1977; Shi et al. 1995) and is an alternative expression of the famous worm-like chain model (Shimada and Yamakawa 1984), which has been proved to be an effective model in describing long DNA fragments (Peters and Maher 2010). In the system of local body coordinates, the governing equations of supercoiled DNA were

$$\frac{d\boldsymbol{F}}{ds} + \boldsymbol{\omega} \times \boldsymbol{F} = 0 \tag{1}$$

$$\frac{d\boldsymbol{M}}{ds} + \boldsymbol{\omega} \times \boldsymbol{M} + \boldsymbol{e}_3 \times \boldsymbol{F} = 0 \tag{2}$$

where $\boldsymbol{F}$ was the internal force, $\boldsymbol{\omega}$ was the torsion rate, $\boldsymbol{M}$ was the inner moment, $\boldsymbol{e}_3$ was the tangent vector and $s$ was the length. The inner moment was $\boldsymbol{M} = (E\frac{\pi a^4}{4}\omega_1, E\frac{\pi a^4}{4}\omega_2, \frac{E}{2(1+\nu)}\frac{\pi a^4}{2}\omega_3)$, where E was the Young's modulus, $\nu$ was the Poisson ratio and $a$ was the radius of the DNA. The magnitude of inner force of DNA was a constant, which gave $F = F\alpha_i \boldsymbol{e_i}$, where $\boldsymbol{\alpha}$ is the direction cosine of the force in the sectional coordinate. The stiff DNA was considered having a modulus twice that of the soft DNA. Same stress boundary conditions were exerted on one end of both the stiff and soft DNA, which resulted an external torsional rate of $1 \pm 0.1\ \text{rad/nm}$ from three directions on the soft and $0.5 \pm 0.1\ \text{rad/nm}$ on the stiff. Each of the DNA fragments (100 nm in length) were set to be randomly oriented and have an initial axial torsional rate $\omega_3{}^0 = 1.8\ rad/n$m. Thus, the torsion rate of the DNA in the local body coordinate could be obtained $\boldsymbol{\omega} = \omega_i \boldsymbol{e_i}$.

2.2 Rebuilding the DNA steric configuration

We denoted the transform tensor between the local body coordinate and the global coordinate as $\boldsymbol{\beta} = \beta_{ij} \boldsymbol{e_i} \boldsymbol{e_j}$, where $\beta_{ij} = \cos(\boldsymbol{e_i}', \boldsymbol{e_j})$. The variation of the three sectional base vectors in the global coordinate was given by

$$\boldsymbol{\beta} \cdot \boldsymbol{\omega} \times \boldsymbol{e}' = \frac{d\boldsymbol{e}'}{ds} \qquad (3)$$

By integrating the tangential vector $\boldsymbol{e_3}'$ through the path, the configuration of DNA yielded

$$\mathbf{x} = \int_0^s \boldsymbol{e_3}' ds \qquad (4)$$

where $\mathbf{x} = x_i \boldsymbol{e_i}$ is the global coordinate of the DNA and $s$ is the length between any position on the DNA and the starting point.

2.3 Stiffness identification

CNN is a revolutionary strategy in image processing for it boosts efficiency and accuracy. With some modifications, VGG (**Fig.1D**), a classical type of CNN, was adopted in this identification task (Simonyan and Zisserman 2014). The VGG architecture consisted of 16 weight layers, which was divided into 5 groups. Batch-norm layers (Bulò et al. 2017) were added between convolutional layers and activation layers to accelerate the training speed and maintain its distribution consistency. In terms of the optimization scheme, ADADELTA (Zeiler 2012) was used to control learning rate and swiftly find the optimal solution. The whole experiment was conducted on the MXNet (Chen et al. 2015), which is an open-source library for deep learning. The trained model was proved robust through several rounds of identifications.

# 3 Results and Discussion

3.1 Simulated images help to establish "stiffness" concept for artificial intelligence.

The trained CNN was tested on another set of 4000 images, which have 2000 stiff DNA images and 2000 soft DNA images. All DNA fragments were set to be randomly oriented. The two orders of supercoiling obtained in our simulation (the second figure in Fig. 1C) agree well with the experimental phenomena (Brady and Fein 1976) and theoretical results (Benham 1977). DNA looping, which has been extensively investigated on (Jeong et al. 2016; Waters and Kim 2013), was also observed in our simulation (the first, second and last figure in Fig. 1A).

The simulated images imported into the artificial neural network in the present study were governed by analytical expression (*i.e.* Kirchhoff elastic rod theory). "Stiffness", the varied parameter in the simulation, is the concept derived in physical models to describe the ability of materials to resist deformation. As for the CNN, a group of the "neurons" (pixels of the feature maps) were activated by the stiff DNA, while another group of neurons were activated by the soft DNA. Some neurons remained silence the whole time. This tendency was verified on these 4000 testing images (**Fig. 1F**). Consequently, the learned features of the CNN had a meaning of the concept "stiffness". The DNA images with random orientation suggested that the learnt "stiffness" were invariable regardless of coordinates.

3.2 The trained CNN can accurately distinguish stiff and soft DNA.

With a closer look at the image set, some of the soft DNAs were much more twisted than the stiff DNAs (**Fig. 1C**). The two kinds of DNAs could be easily distinguished (**Fig. 1H**). However, a more challenging task involved two very similar samples (**Fig. 1C**). Surprisingly, they could also be accurately classified with rather high certainty (**Fig. 1H**). The trained convolutional kernels of the first layer (**Fig. 1E**) could accurately extract the features of DNA images (**Fig. 1G**). The overall prediction accuracy reached 99.85%.

The stiffness of DNA could also be calculated by computing the tangent-tangent cross-correlation along the DNA contour (Cassina, Manghi, Salerno, Tempestini, Iadarola, Nardo, Brioschi and Mantegazza 2016). This semi-automated approach identifies the mechanical properties of DNA based on WLC (worm-like chain) model and from certain geometrical information (*e.g.*, end-to-end distance of DNA segment), the accuracy of which is limited by the WLC model. On the contrary, the advantages of CNN lie in its sensitivity to dimensionless features and complete automation. By convoluting the figure with kernels, the CNN-system identifies the stiffness of DNA from all the geometrical features and stiffness-relating information, whose accuracy is not limited by physical models. Thus, the CNN-system has the potential to realize higher accuracy of identification. Besides, the identification by CNN is completely automated, which is better than the semi-automated approach as it saves manpower. However, the robustness Cassina's semi-automated calculation is obvious, while that of the CNN in the physical situation should be further confirmed. When applied to the real situation, we suggest that governing equations (*i.e.* Kirchhoff theory or worm-like chain model) could be incorporated into the loss function of neuron network when adopted in the real situation so that the neuron network could be physical informed.

Very-long and filament-like structures can be found within the cell nucleus and its environment, among which are DNA (Lyubchenko and Shlyakhtenko 1997), nucleoskeleton (Hozák et al. 1995), nuclear basket (Arlucea et al. 1998), cytoskeleton (Fletcher and Mullins 2010) , and *etc.* These structures, as a result of their length, mostly possess curved and torsional steric configuration. A crucial problem related to large-scale *in-vivo* nuclear research is how to automatically identify and classify these micro-filaments. Both static and dynamic behavior of DNA could be recorded by atom force microscopy (AFM) (Lyubchenko and Shlyakhtenko 2016). For example, DNAs in the presence of alkaline earth metal ions could be identified by AFM (Zheng et al. 2003). With advanced tips, the resolution of AFM reaches 1.2 nm which makes it possible to resolve the double-helix structure of DNA (Li et al. 1999). Time-Lapse AFM helped to observe the unwrapping process of nucleosome during which the DNA could be identified (Shlyakhtenko et al. 2009), and the substrate surface could be modified to mimic intracellular surface *in vivo* (Lyubchenko 2014). Combining the AFM with the proposed approach, real-time, fast and non-expert diagnostics of DNAs are potential in future.

**4 Conclusion**

This study proved the efficiency of deep learning on the identification of supercoiled DNA stiffness, which may be a promising approach for the point-of-care diagnostics. It also provided a framework to implant the physical model into the artificial intelligent model.

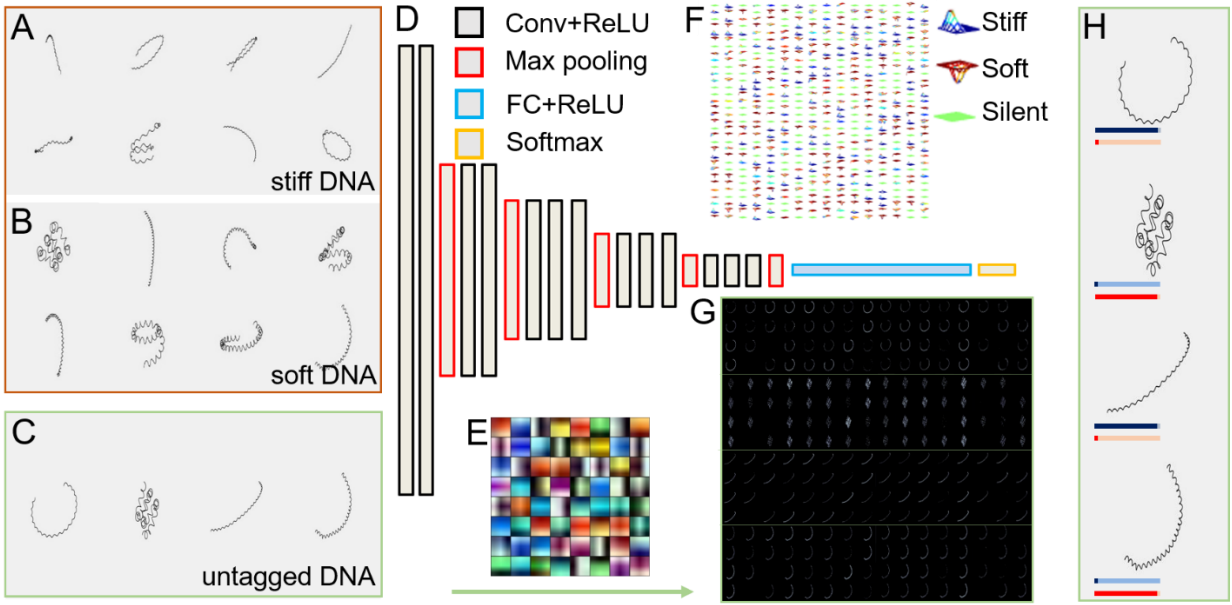

**Figure 1. The framework of training and predicting of DNA fragments with a CNN (convolutional neural network).** A & B. labelled DNA as the training set; C. untagged DNA to be classified; D. a classical structure (VGG16) of CNN. Conv: convolutional layer; ReLU: rectified linear unit; FC: fully connected layer; E. 64 trained convolutional kernels (3×3 feature extractor) in the bottom layer, indicating the learned features of the DNA graphs; F. Average activity of "artificial neurons" in the top layer. The top layer consisted of 512 feature maps (7 × 7 pixels). G. 64 feature maps of the DNA features for four typical DNA fragments in the bottom layer of the CNN. The lightened region in the graph meant the information in this region was extracted by the corresponding channels. H. predicted results of the untagged DNA. The bar shows the certainty for predicted labels of two types: blue bar stands for stiff DNA and red bar stands for soft DNA.

**Acknowledgments**

This work was financially supported by the New Faculty Foundation of NUAA (1001-YAH19016), by the foundation of "Jiangsu Provincial Key Laboratory of Bionic Functional Materials" (1001-XCA1816310), and by the Foundation for the Priority Academic Program Development of Jiangsu Higher Education Institutions.

**Competing interests**

The authors declare no competing interests.

**References**

ALIPANAHI, B., A. DELONG, M. T. WEIRAUCH AND B. J. FREY Predicting the sequence specificities of DNA- and RNA-binding proteins by deep learning. Nature Biotechnology, 2015, 33(8), 831.

ANGERMUELLER, C., T. PäRNAMAA, L. PARTS AND O. STEGLE Deep learning for computational biology. Molecular Systems Biology, 2016, 12(7), 878.

ARLUCEA, J., R. ANDRADE, R. ALONSO AND J. ARéCHAGA The Nuclear Basket of the Nuclear Pore Complex Is Part of a Higher-Order Filamentous Network That Is Related to Chromatin. Journal of Structural Biology, 1998, 124(1), 51-58.

BAIR, E. AND R. TIBSHIRANI Machine learning methods applied to DNA microarray data can improve the diagnosis of cancer. Acm Sigkdd Explorations Newsletter, 2003, 5(2), 48-55.

BENHAM, C. J. Elastic model of supercoiling. Proceedings of the National Academy of Sciences of the United States of America, 1977, 74(6), 2397-2401.

BENHAM, C. J. Geometry and mechanics of DNA superhelicity. Biopolymers, 2010, 22(11), 2477-2495.

BOYNTON, K. A., I. C. SUMMERHAYES, D. A. AHLQUIST AND A. P. SHUBER DNA integrity as a potential marker for stool-based detection of colorectal cancer. Clinical Chemistry, 2003, 49(7), 1058.

BRADY, G. W. AND D. B. FEIN X-ray diffraction studies of circular superhelical DNA at 300-10,000-A resolution. Nature, 1976, 264(5583), 231.

BULò, S. R., L. PORZI AND P. KONTSCHIEDER In-Place Activated BatchNorm for Memory-Optimized Training of DNNs 2017.

CASSINA, V., M. MANGHI, D. SALERNO, A. TEMPESTINI, et al. Effects of cytosine methylation on DNA morphology: An atomic force microscopy study. Biochim Biophys Acta, Jan 2016, 1860(1 Pt A), 1-7.

CHEN, T., M. LI, Y. LI, M. LIN, et al. MXNet: A Flexible and Efficient Machine Learning Library for Heterogeneous Distributed Systems. Statistics, 2015.

CUMMINGS, B. B., J. L. MARSHALL, T. TUKIAINEN, M. LEK, et al. Improving genetic diagnosis in Mendelian disease with transcriptome sequencing. Science Translational Medicine, 2017, 9(386), eaal5209.

FLETCHER, D. A. AND R. D. MULLINS Cell mechanics and the cytoskeleton. Nature, 2010, 463(7280), 485-492.

GOOTENBERG, J. S., O. O. ABUDAYYEH, J. W. LEE, P. ESSLETZBICHLER, et al. Nucleic acid detection with CRISPR-Cas13a/C2c2. Science, 2017, 356(6336), 438-442.

HOZáK, P., A. M. SASSEVILLE, Y. RAYMOND AND P. R. COOK Lamin proteins form an internal nucleoskeleton as well as a peripheral lamina in human cells. Journal of Cell Science, 1995, 108 ( Pt 2)(Pt 2), 635-644.

HUANG, J., T. SCHLICK AND A. VOLOGODSKII Dynamics of site juxtaposition in supercoiled DNA. Proceedings of the National Academy of Sciences of the United States of America, 2001, 98(3), 968-973.

JEONG, J., T. T. LE AND H. D. KIM Single-molecule fluorescence studies on DNA looping. Methods, Aug 1 2016, 105, 34-43.

JIAN, H., T. SCHLICK AND A. VOLOGODSKII Internal motion of supercoiled DNA: brownian dynamics simulations of site juxtaposition. Journal of Molecular Biology, 1998, 284(2), 287.
KELLEY, D. R., J. SNOEK AND J. RINN Basset: Learning the regulatory code of the accessible genome with deep convolutional neural networks. Genome Research, 2016, 26(7), 990.
KREMER, L. S., D. M. BADER, C. MERTES, R. KOPAJTICH, et al. Genetic diagnosis of Mendelian disorders via RNA sequencing. Nature Communications, 2017, 8, 15824.
KUHN, L., H. Y. KIM, J. WALTER, D. M. THEA, et al. HIV-1 concentrations in human breast milk before and after weaning. Science Translational Medicine, 2013, 5(181), 181ra151.
LI, J., A. M. CASSELL AND H. DAI Carbon nanotubes as AFM tips: Measuring DNA molecules at the liquid/solid interface. Surface & Interface Analysis, 1999, 28(1), 8-11.
LYUBCHENKO, Y. L. Nanoscale nucleosome dynamics assessed with time-lapse AFM. Biophysical Reviews, 2014, 6(2), 181-190.
LYUBCHENKO, Y. L. AND L. S. SHLYAKHTENKO Visualization of Supercoiled DNA with Atomic Force Microscopy in situ. Proceedings of the National Academy of Sciences of the United States of America, 1997, 94(2), 496-501.
LYUBCHENKO, Y. L. AND L. S. SHLYAKHTENKO Imaging of DNA and Protein−DNA Complexes with Atomic Force Microscopy. Critical Reviews in Eukaryotic Gene Expression, 2016, 26(1), 63.
MOINOVA, H. R., T. LAFRAMBOISE, J. D. LUTTERBAUGH, A. K. CHANDAR, et al. Identifying DNA methylation biomarkers for non-endoscopic detection of Barrett's esophagus. Science Translational Medicine, 2018, 10(424), eaao5848.
OU, C. Y., S. KWOK, S. W. MITCHELL, D. H. MACK, et al. DNA amplification for direct detection of HIV-1 in DNA of peripheral blood mononuclear cells. Science, 1988, 239(4837), 295.
PARK, S. J., T. A. TATON AND C. A. MIRKIN Array-based electrical detection of DNA with nanoparticle probes. Science, 2002, 295(5559), 1503-1506.
PETERS, J. P., 3RD AND L. J. MAHER DNA curvature and flexibility in vitro and in vivo. Q Rev Biophys, Feb 2010, 43(1), 23-63.
SHI, Y., A. E. BOROVIK AND J. E. HEARST Elastic rod model incorporating shear and extension, generalized nonlinear Schrödinger equations, and novel closed-form solutions for supercoiled DNA. Journal of Chemical Physics, 1995, 103(8), 3166-3183.
SHIMADA, J. AND H. YAMAKAWA Ring-closure probabilities for twisted wormlike chains. Application to DNA. Macromolecules, 1984, 17(4), 689-698.
SHLYAKHTENKO, L. S., A. Y. LUSHNIKOV AND Y. L. LYUBCHENKO Dynamics of nucleosomes revealed by time-lapse atomic force microscopy. Biochemistry, 2009, 48(33), 7842-7848.
SIMONYAN, K. AND A. ZISSERMAN Very Deep Convolutional Networks for Large-Scale Image Recognition. Computer Science, 2014.
SKOLNICK, J. AND M. FIXMAN Electrostatic Persistence Length of a Wormlike Polyelectrolyte. Macromolecules, 2010, 10(5), 944-948.
TRAVERS, A. A. The Structural Basis of DNA Flexibility. Philosophical Transactions Mathematical Physical & Engineering Sciences, 2004, 362(1820), 1423-1438.
UHLER, C. AND G. V. SHIVASHANKAR Regulation of genome organization and gene expression by nuclear mechanotransduction. Nature Reviews Molecular Cell Biology, 2017, 18(12), 717.
WATERS, J. T. AND H. D. KIM Equilibrium Statistics of a Surface-Pinned Semiflexible Polymer. Macromolecules, 2013, 46(16), 6659-6666.
YUAN, C., H. CHEN, X. W. LOU AND L. A. ARCHER DNA bending stiffness on small length scales. Phys Rev Lett, Jan 11 2008, 100(1), 018102.
ZEILER, M. D. ADADELTA: An Adaptive Learning Rate Method. Computer Science, 2012.
ZHENG, J., L. I. ZHUANG, W. U. AIGUO AND H. ZHOU AFM studies of DNA structures on mica in the presence of alkaline earth metal ions. Biophysical Chemistry, 2003, 104(1), 37-43.